# Understanding the compromise between Skyline and Ranking queries

Marco Tonnarelli

Politecnico di Milano
Milan, Italy
marco.tonnarelli@mail.polimi.it

**Abstract**

Skyline and Ranking queries have gained great popularity in the recent years. These two techniques are crucial for multi-criteria decision support applications, which are now more popular than ever before. Skyline and Ranking queries are, however, affected by well-known limitations. In the past recent years, the database community provided numerous studies in this field with the aim to overcome the weaknesses of these two approaches. This survey introduces the reader to Skyline and Ranking queries, explaining the concepts on which they are based, with the intent to present the compromise between the two techniques: flexible skylines.

## 1 Introduction

Suppose you want to buy a used motorcycle through an online retailer, and you are looking for an inexpensive one with low mileage. Achieving this goal might be difficult, as the two attributes conflict with each other: a lower mileage motorcycle tends to have a higher price. For example, a motorcycle with 10000 miles and a price of $5000 is perhaps *preferable to* a motorcycle with 11000 miles and a $7000 price. The online retailer's search engine needs an effective and efficient ranking method to show the most *interesting* results *for you. Skyline* and *top-k* queries arise in this kind of scenario.

The two techniques, which have been extensively studied in the past decades [1], are based on one fundamental concept: identify records of preference in a multi-objective setting. Nonetheless, the two paradigms rely on very different and well-studied approaches [1, 2, 3, 4]: skyline queries are based on the concept of *dominance*, whereas ranking queries rely on the concept of *utility function*, which is defined over the records' attributes. These two techniques differ in terms of performance, output size and control of cardinality; furthermore, each one has benefits and drawbacks.

In subsections 1.1 and 1.2 we will describe the main characteristics of, respectively, skyline queries and ranking queries, highlighting the differences between them. Subsequently, we will discuss how it is possible to overcome the main limitations of the two paradigms by introducing recently developed methods, i.e., the *restricted* or *flexible* skylines. The main differences of *skyline* and *ranking* queries are summarized in Table 1.

|  | Simplicity | Control of cardinality | Overall view of interesting results | Input flexibility |
|---|---|---|---|---|
| Ranking Queries | No | Yes | No | Yes |
| Skyline Queries | Yes | No | Yes | No |

Table 1: Pros and cons of ranking and skyline queries.

## 1.1 Skyline Queries

Computing a skyline query problem means solving the maximum vector problem [2, 3], which consists of finding the records (points) that are not dominated by any others. A record dominates another record if all its attributes are as good or better, and at least one is better. The dominance relation is transitive, i.e., if *a* dominates *b* and *b* dominates *c*, then *a* dominates *c*. The *skyline* includes all the records that are not dominated.

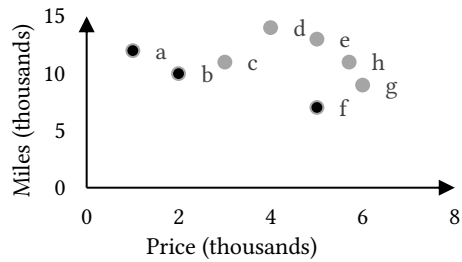

Figure 1: used motorcycles example.

Figure 1 shows an instance of our introductory example, where points *a*, *b* and *f* are dominated by no other point, i.e., *a*, *b* and *f* are the most interesting points in the dataset. The computation of *skyline* queries originally [2] assumed that all the data were present in main memory, but nowadays this assumption is no longer possible since the size of datasets is enormously increasing. Thus, being data stored on secondary memory, in the past years many algorithms [5] were proposed to speed-up computation, by preprocessing data: Block-Nested-Loop, Divide and Conquer, Bitmap, B-Trees and R-Trees. The main objective of these procedures is computing, at a given time, only a subset of the original dataset on the main memory rather than directly working on its entirety.

The performance of these algorithms is strictly related to the size of the dataset and the number of dimensions (attributes) they are working on; as the number of dimensions increases, the problem's complexity increases [1, 4].

The absence of a utility function makes the skyline operator simpler with regard to ranking queries but, consequently, this simplicity leads to the lack of control over the cardinality of the output. Indeed, the optimization achieved by the aforementioned algorithms is limited by the size of the datasets: the larger the dataset, the lower the probability that a record dominates another record and thus the scale of the obtained result is unpredictable [4, 5]. The algorithms proposed in these studies are based on the maximum vector problem but there are other possibilities that solve strictly related problems, i.e., the *contour problem* [6], *multi-objective optimization* using linear programming [7] and the computation of the *convex hull* of a dataset [3] (which is a subset of the skyline).

## 1.2 Ranking Queries

*Ranking* queries, also known as top-*k* queries, aim to retrieve the *k* best answers from a dataset, i.e., the most interesting, relevant, or important instances for the user. Ranking means ordering the results based on the *relevance* they have to the query, and this is possible by introducing the concept of *utility* *function* (we also refer to it as *scoring function*). The most common approach [1, 9, 10] consists of assigning a score to all the records in the dataset, by means of the scoring function, and then identifying the best scoring ones. Assigning a score to a record means making a valuation over the attributes and, typically, the scoring function is defined as the aggregation of partial scores over each attribute [1]. However, the scoring function might be implemented in several forms as different applications require exploiting features that are peculiar to the specific scenario in which they are employed: generic function, monotone function, and no ranking function (in this case the problem overlaps the skyline paradigm) are the typical [1] classifications.

The utility function is defined as follows: $f(t) = \sum_{i=1}^{d} p_i * t_i$ , where $i$ denotes the $i$-th element, $d$ the number of attributes $t$ and $p_i$ the weight assigned to an attribute.

| Product | Price | Miles |
|---|---|---|
| N1 | 4000 | 13000 |
| N2 | 1000 | 9000 |
| N3 | 3500 | 10000 |
| N4 | 2800 | 8000 |
| N5 | 3000 | 11000 |

| User | $p_1$ | $p_2$ | Top-2 result |
|---|---|---|---|
| 1 | 0.74 | 0.26 | N2, N4 |
| 2 | 0.10 | 0.90 | N4, N2 |

Table 1.1: e-commerce motorcycles example.     Table 1.2: top-2 results.

Tables 1.1 and 1.2 show an instance of our running example, where each user has a customized preference vector and, therefore, the top results of the query are different: in this specific case, the returned records are the same, but the ranking is different.

With this definition, each record has a score value (*already assigned*), and the naïve approach to compute the top-k elements consists of *sorting* the records by the *scoring function* values [1, 8]. However, this approach becomes quickly computationally intensive as the number of objects in the dataset increases. Furthermore, we are assuming that the score assigned to each element in the database is computed before the query is evaluated. Many algorithms [1] have been developed to efficiently process ranking queries and are classified on the basis of different design levels: according to the query model they assume, the method through which they access data, the level on which they are implemented, the uncertainty over data and query model or the restriction imposed to the ranking function.

**1.3 Skyline and Ranking Queries Shortcomings**

As mentioned in the previous paragraphs, skyline and top-k queries are affected by several limitations, that we summarize in this subsection. The main issue with skyline queries is related to the fact that the user is unable to predict the size of the output of the skyline, and the control over the output size is crucial for the quality of the decision the user takes. This is known by Hick-Hyman Law [19, 20], which states that more choices the user faces, the longer it will take them to make a good decision. Another issue affecting skyline queries is to be traced back to the lack of customization: indeed, the result of a skyline is the same for every user, and this is true because of their invariance to dimensional scaling (assigning a weight to each dimension does not change the result).

These two issues, apparently, do not exist if we consider ranking queries, as they are capable of customization of the results and, also, they give control over output size. Nonetheless, also ranking queries have issues, and the most relevant is the following. The user preferences' vector, i.e., the scoring function, is assumed to be provided by the user (or, alternatively, mined by previous user's decisions). This assumption is unrealistic, as a minor change in the weights could significantly alter the top-k results' set [1, 8]. Thus, in modern applications, preference weights are used as a guide, rather than the exact representation of the users' preference [8, 10]. Consequently, another drawback of top-k queries is the complexity of implementation, as it becomes necessary to deal with uncertainty.

In past recent years, several studies have been conducted with the objective of overcoming these drawbacks, mainly by finding a trade-off between the two paradigms, to preserve the advantages of both techniques. In this paper we describe the most significant techniques: F-skyline, ε-skyline, k-regret query, trade-off skyline, UTK, ORD and ORU.

## 2 ε-Skyline

The *ε-skyline* operator [8] is the first one we analyze in this survey: output customization and control over result's cardinality are the focus of this paradigm. To achieve these results, ε-skyline applies the concepts of scoring function and output size control, provided by top-k queries, to the traditional skyline operator. This is made possible by the introduction of a new flexible relation: *ε-dominance*. Before the introduction of this new paradigm, already existing methods allowed increasing or decreasing the number of returned objects but doing both was not possible: ε-skyline allows overcoming also this drawback.

The ε-dominance relation is defined as follows: given a set of tuples with d-attributes and a set of weights $W = \{w_i \mid i \in [1, d], 0 < w_i \leq 1\}$, and a constant $\varepsilon \in [-1,1]$, for any two tuples $t_1$ and $t_2$ is said that $t_1$ *ε-dominates* $t_2$, denoted as $t_1 \prec_\varepsilon t_2$, if $\forall_i \in [1, d], t_1[i] \cdot w_i \leq t_2[i] \cdot w_i + \varepsilon$, and $\exists j \in [1, d], t_1[j] < t_2[j]$.

The *ε-Skyline* returns all the records in the database that are not *ε-dominated* other records. The ε parameter adjusts the size of the set returned by the query: the smaller ε, the smaller the set, and vice versa. If ε is equal to zero, it means that we are dealing with a conventional skyline. This parameter can be interpreted as a similarity index between two records: if one record is *slightly better* than another, the former is returned to the user.

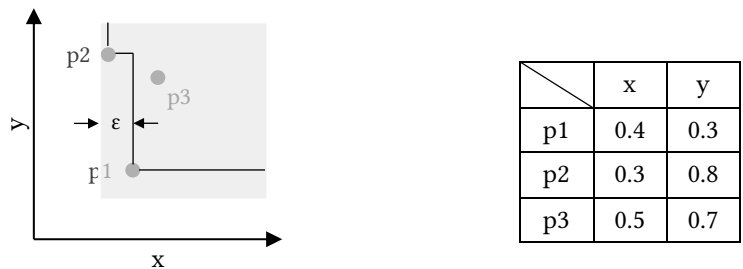

Figure 2: two-dimensional dataset example.

Figure 2 shows an example of a two-dimensional dataset. The dark line that connects the points depicts the original skyline dominance region, while the light grey area represents the ε-dominance region of record *p1* when ε is set to 0.1. This simple example clearly demonstrates how the result set vary as the dominance region changes: the original skyline result comprises records *p1* and *p2*, while the ε-skyline returns only record *p1* since *p2* is ε-dominated by *p1*.

### 2.1 ε-Skyline Computation

It is not possible to compute ε-skyline with traditional skyline algorithms, due to the lack of strict transitivity and asymmetry properties. However, it is possible to extend the Sort Filter Skyline progressive algorithm: `ε-SFS` Algorithm can compute the new operator, at the cost of presorting all the objects in the database.

A more effective way to compute this operator consists of the application of the Index-based Filter Refinement (IFR) algorithm. `IFR` algorithm exploits a data-partitioning spatial index on the dataset to achieve more pruning power. With spatial index, tuples are partitioned in regions and the algorithm proceeds, for each record, by checking which regions are ε-dominated. Regions that are strictly dominated by a record can be safely pruned, whereas regions that are not strictly dominated by a given record *r*, but reside in the ε-dominant region of *r*, might *affect r*: any object in the affecting region of *r* ε-dominates *r*. If there is at least a record in the affecting region, the algorithm discards *r*. Applying this procedure recursively leads to the evaluation of the resulting set of the ε-Skyline.

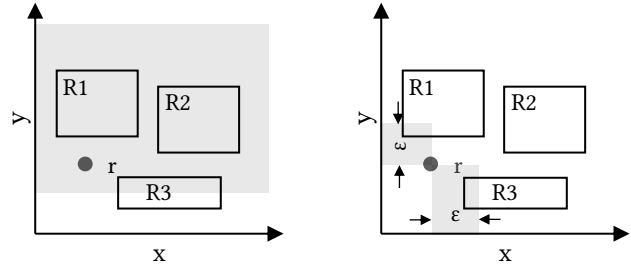

Figure 3: Spatial index example.

Figure 3 shows an example of regions (minimum bounding rectangles of objects) and how they are managed by the algorithm (the shaded region in figure 3.1 represent the ε-dominant region of the point *r*). In figure 3.1 is visible that R2 can be safely pruned as it is dominated by the point. It is not possible to prune R1 as some records inside it might ε-dominate the considered point (R1 affects the point). Figure 3.2 displays the affecting region of the point. Finally, region R3 has at least one record ε-dominating the point because it has an edge in the affecting region: so, the point can be directly eliminated.

The main issue with this paradigm lies in the choice of the right $\varepsilon$ value, as it changes dramatically the output size and the `IFR` Algorithm is very sensitive to it.

## 3 F-Skyline

F-skylines [21], also known as *restricted skylines* [9], are the second paradigm we analyze. This operator adds user preference to traditional skylines by considering arbitrary families of scoring functions: at the basis of the F-skylines lies the concept of $\mathcal{F}$-*dominance*. Another benefit of this approach consists of the reduction of the output size. $\mathcal{F}$-dominance is defined as follows: an object $t$ $\mathcal{F}$-dominates another object $r$ when $t$ is better than or equal to $r$ based on all the scoring functions in the family $\mathcal{F}$, and strictly better for at least one function in $\mathcal{F}$.

There are two F-Skyline operators: ND, which describes the set of *non-$\mathcal{F}$-dominated* tuples, and PO, which describes the set of *potentially optimal* tuples, i.e., the best tuples according to a function in the family $\mathcal{F}$. The relation between F-skylines and traditional skylines is defined as follows: PO is a subset of ND, which is a subset of traditional skylines if we consider a generic family $\mathcal{F}$ of functions; otherwise, if we consider the set of monotone scoring functions $\mathcal{M}$, PO and ND sets coincide with skylines.

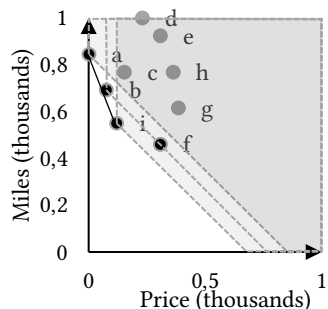

Figure 4: $\mathcal{F}$-dominance regions of tuples *a, b* and *i* within motorcycle dataset (normalized in [0,1]), with f=$w_1$x+$w_2$y such that $w_1 \geq w_2$.

The computation of the F-skyline query is related to the evaluation of the $\mathcal{F}$-dominance region, namely DR(t; F), which is defined for a tuple $t$, with respect to a set of monotone scoring functions, as the set of all points that are $\mathcal{F}$-dominated by $t$ in the $d$-dimensional space [0,1].

Figure 4 shows an application of $\mathcal{F}$-dominance to our running example, i.e., buying a used motorcycle though an online retailer, with the addition of new point $i$. In the figure, data are normalized in region [0,1]. If we consider a family of functions in the form of $w_1 x + w_2 y$ such that $w_1 \geq w_2$, the $\mathcal{F}$-dominance region of tuples $a$, $b$ and $i$ is displayed in grey. The normal skyline set is made by points $a$, $b$, $f$ and $i$; however, the ND set is made only by points $a$, $b$ and $i$,, which is different from the PO set, as it is constituted by records $a$ and $i$: $PO \subseteq ND \subseteq Sky$. This example shows how these operators can reduce the resulting set size: in fact, the considered dataset is small, but as the dataset increases, the improvement might increase.

### 3.1 F-Skyline Computation

It is possible to compute the ND set in several different ways, and the main distinction between the available algorithms is made on whether the technique is divided in phases, i.e., the computation of ND set comes after the computation of skyline set otherwise contemporarily, or the algorithm contemplates presorting the dataset or not. In general, sorted strategies perform better then unsorted strategies.

`SLP2` and `SVE2` are examples of two-phases algorithms, which scan the sorted tuples and populate the current window of ND tuples among those in the skyline set. Being the dataset sorted, no tuples will be removed from ND. With respect to one-phase algorithms, `SVE1F` interleaves dominance and $\mathcal{F}$-dominance tests, and thus it performs a single scan over ND, by checking first $\mathcal{F}$-dominance.

PO operator works by starting from the tuples in ND set. Algorithm `POND` is able to reduce the set of potentially optimal candidate tuples by the application of some heuristics: by iteratively increasing the set (size $n$) of convex combination of tuples, and after reverse ordering (worst tuples are most likely to be $\mathcal{F}$-dominated) of the PO tuples (which set is initially defined as ND), if a tuple $s$ $\mathcal{F}$-dominates another tuple $t$ then $s$ is a convex combination of the first $n$ tuples in PO; thus, we prune these tuples. This happens until the size of set of size $n$ is greater than the size of PO set.

Although these paradigms add user preference and reduction of the output size, it is still not possible to predict the size of the output [10].

## 4 Uncertain Top-k Query

UTK query [22] is a technique born with the intent to consider the preference vector $w$ provided by the user only as an estimate. UTK query deals with this issue by expanding the weight vector into a region and reporting all top-k sets to the user. If we consider our running example, a useful feature introduced by this paradigm is the following. Suppose you received a set of top-k recommendations of motorcycles which satisfy your preference: you might want to ask for additional suggestions on top of the previous ones; so, we can expand the weight vector into a region in order to retrieve additional options similar to the first ones. Although we are adding uncertainty on the wight vector, there is nothing probabilistic in UTK: the output is exact, and the records' score are correlated as they vary together with $w$. UTK is related to the *k-skyband* [23], which is a generalization of the traditional skyline: it is defined as a set of points which are dominated, in the original skyline meaning, by at most k-1 others, and it is a superset of the records that appear in a top-k result for any weight vector.

There are two versions of UTK: UTK$_1$, which returns the minimal set of records that rank as top-k when $w$ resides in R (where R is defined as a region in the preference domain) and UTK$_2$, which returns the minimal set of records that rank as top-k for all the possible vectors in R.

UTK queries are based on one important notion: *r-dominance*: record $a$ r-dominates record $b$ if $a$ scores higher or equal than $b$ for any weight vector in R, and score strictly higher than $b$ for at least one weight vector in R. Two records are *r-incomparable* if the hyperplane U($a$)=U($b$) (where U($x$) denotes the utility score) lies among region R, i.e., the hyperplane "cuts" R. Otherwise, a record might dominate another record or might be dominated.

For instance, we consider now our motorcycles running example. We have a 2-dimensional dataset D and, if we assume preference weights $w_i \in (0,1)$ and $\sum_{i=1}^{2} w_i = 1$, we can reduce to a 1-dimensional preference domain, as we can derive $w_2 = 1 - w_1$. So, we can plot the scores of the records as functions of $w_1$. Figure 5 shows the score of records *b, c, d, e, g*, denoted with $r^*$. It is displayed also region R (in bold, between 0.3 and 0.7): we can see, for k=2, the first two levels of dominance, i.e., the highest scoring records; in particular, the short-dashed line identifies level 1 and the long-dashed line identifies level 2.

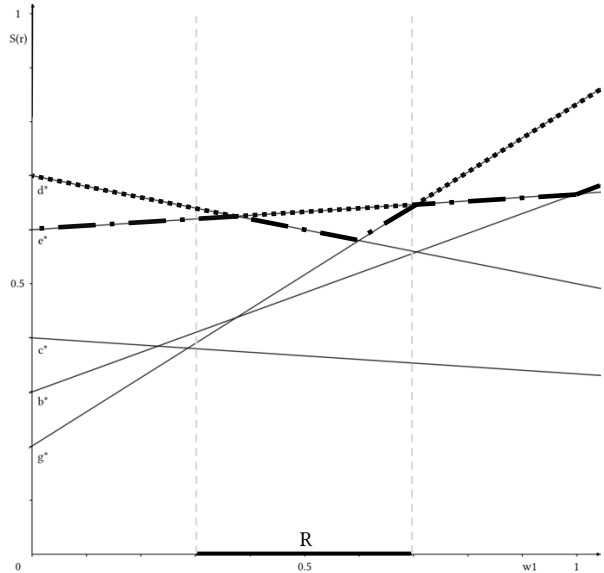

Figure 5: 1-dimensional preference domain and region R of records b, c, d, e, g of our running example.

### 4.1 UTK Computation

There are different procedures to compute the two UTK query versions, since $UTK_2$ produces a more detailed output than $UTK_1$: for instance, R-Skyband Algorithm (`RSA`) is used to compute $UTK_1$ problems, while Joint Arrangement Algorithm (`JAA`) deals with $UTK_2$ problem.

`RSA` works by first computing the k-skyband set, which is a superset of $UTK_1$ result, together with the *r-dominance graph* to keep track of all the pair-wise r-dominance relationships between records. Then, in a refinement step, each record in the r-skyband is considered in decreasing order of the *r-dominance count*, i.e., the number of records that r-dominate a record *t*, to confirm if a tuple belongs to the $UTK_1$ result, by analyzing the r-dominance graph.

`JAA` is similar to `RSA`, since it builds the r-dominance graph starting from the k-skyband. However, during the refinement step, records are not taken into account individually, instead it is considered a partitioning of R for all the candidates.

## 5 k-Regret Query

K-representative regret minimization queries [15], namely *k-regret*, are another approach which is based on the use of the concept of utility function, but without assuming that the preference is provided by the users, as opposed to top-k queries. The purpose of this technique is to capture how *unhappy* the user is when they see *k* representative tuples instead of the entire dataset. The happiness of the user is measured as x%, which denotes the percentage of utility they get from the best tuple in the list, at least *x%* of the utility they get from best tuple in the entire dataset. This operator provides a set of at most k tuples without asking to the user the utility function: instead, it tries to output the set that makes every user *x%* happy. This is possible by minimizing the *maximum regret ratio*, which is defined as:

$$rr_D(S,F) = sup_{f \in F} rr_D(S,f) = sup_{f \in F} 1 - \frac{max_{p \in S} f(p)}{max_{p \in D} f(p)},$$

$$\text{where } rr_D(S,f) = \frac{r_D(S,f)}{gain(D,f)} \text{ is the regret ratio,}$$

$$\text{the regret is } r_D(S,f) = gain(D,f) - gain(S,f),$$

$$\text{and } gain(S,f) = max_{p \in S} f(p)$$

which is the worst possible regret for any user with a utility function in F[1], and it is defined over input set D and output set S ($S \subseteq D$). The *regret* is one minus the happiness of the user. Another important concept is *gain,* which is the maximum utility derived from the subset of the k-Regret query. This operator is *scale invariant* with respect to the attributes and *stable,* i.e., adding not relevant tuples to the database does not change the result. The maximum regret ratio can be bounded in terms of *k* representative tuples and dimensions *d.*

Given that the utility function is not provided by the user, the evaluation is made by considering a broad family F of function and maximizing over the worst.

|   | Price | Mileage | $f_{0.3,0.7}(p)$ | $f_{0.6,0.4}(p)$ | $f_{0.2,0.8}(p)$ |
|---|---|---|---|---|---|
| b | 2 | 10 | 7.60 | 5.20 | 8.40 |
| c | 3 | 11 | 8.60 | 6.20 | 9.40 |
| d | 4 | 14 | 11.00 | 8.00 | 12.00 |
| e | 5 | 13 | 10.60 | 8.20 | 11.40 |
| g | 6 | 9 | 8.10 | 7.20 | 8.40 |
| h | 5.7 | 12 | 10.11 | 8.22 | 10.74 |
| f | 5 | 7 | 6.40 | 5.80 | 6.60 |
| a | 1 | 12 | 8.7 | 5.40 | 9.80 |
| Regret ratio of tuples {h, e} | | | $1 - \frac{10.60}{11.00} = 3.64\%$ | $1 - \frac{8.22}{8.22} = 0\%$ | $1 - \frac{11.40}{12.00} = 5\%$ |

Table 2: motorcycle dataset, utilities, and regret ratio (units in thousands).

For example, Table 2 shows an instance of our running example, where we consider a set of utility functions. For each record, it is indicated the respective utility score. By applying the definition of regret ratio, given sets D = {a, b, c, d, e, f, g, h} and S = {h, e}, we evaluate the ratio for each utility. The k-regret query retrieves a small set of tuples from the database such that the utility of the favorite among these tuples is a small fraction, smaller than the utility of the favorite among all the tuples in the dataset. The results of this query might be presented in the home page of the online retailer's website, as the query minimized the regret of all users: this means that the result set contains tuples which will most probably attract many users. In this example, the maximum regret ratio is computed by
$rr_D(S,F) = sup_{f \in F} rr_D(S,f) = max\{3.64\%, 0\%, 5\%\} = 5\%.$

## 5.1 k-Regret Computation

Computing the optimal solution of a regret minimization set problem is NP-hard if some conditions on dimensionality of dataset are met [18]. There exist several solutions to compute this kind of queries [15, 16, 17] either heuristic or theoretical algorithms. Among all the non-theoretical techniques and considering only algorithms that do not require preprocessing, `GREEDY` and `MAX-DOM-GREEDY` are the best overall techniques, given that they give solutions with maximum regret ratio much lower than the theoretical bound [15].

# 6 Trade-Off Skyline

As we previously described, regular skyline queries are based on the dominance concept. Based on this notion, *trade-off skylines* [11] add a customizable data filter on top of it. This is achieved by the fact that the user can declare some preferences over the attributes, or it might be elicited with some heuristics [13]. This preference over the attributes could be numerical but also qualitative (in this case the preference is modeled by weak order [14]). The main objective of this technique is to recreate in the query process the user's everyday decision processes. The trade-off is intended as the focus of the user over a subset of the available attributes: it represents in a qualitative manner how much the user is willing to sacrifice some dimensions to obtain higher performance over other attributes. The trade-off is a relationship between two tuples, denoted as $t := (x \triangleright y)$, defined over a set of attributes μ, i.e., a set of indexes {1, … n}, where $n$ is the total number of attributes. Each trade-off induces new dominance relationship between database items: given a set of trade-offs T, the dominance relation is denoted with $>_T$, i.e., a record dominates another record with respect to the trade-off set T.

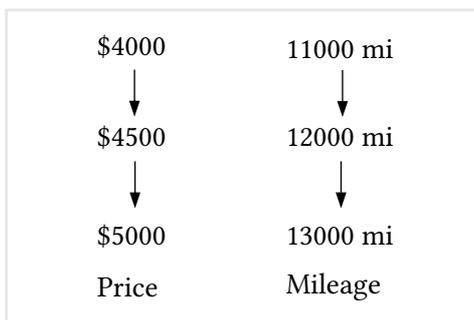 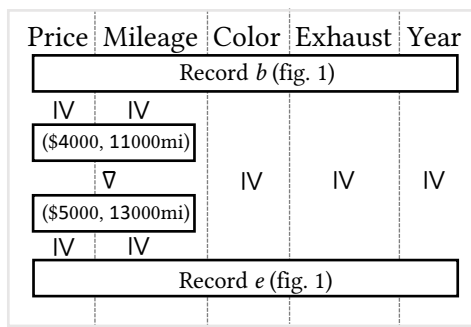

Figure 6.1: preferences within our running example.

Figure 6.2: trade-off domination.

For instance, we consider our running example. Suppose to have more dimensions, i.e., more attributes, such as the color, installed aftermarket exhaust and year of registration. Each user might have different preferences over the attributes, and, for example (Figure 6.1), one user might consider more relevant price and mileage within all the others. In particular, a lower price and a lower mileage are preferable: considering the previous definition, the trade-off is expressed by (($4000, 11000mi) ▷ ($5000, 13000mi)). If we consider Figure 1 and the trade-off domination relationship previously defined, record *b* dominates record *e*. The attributes Price and Mileage constitute the μ whereas Color, Exhaust and Year represent the attributes excluded by the trade-off.

## 6.1 Trade-Off Skyline Computation

The simplest way to computer trade-off queries consists of computing first the traditional skyline and then add the new constraints provided by the trade-off set and apply them to the result of the skyline: this is possible because the *trade-off skyline* is a subset of the regular skyline. However, the trade-off check is computationally more expensive and, thus, interleaving these checks is computationally prohibitive. Basically, the Basic Trade-off Algorithm works by first computing the regular skyline and then, checks if the domination hold for each trade-off *t* in set T. As previously mentioned, this is not efficient.

A more efficient way to computed trade-off queries involve the representation of the trade-off sequences by incrementally building a tree structure, namely *TTree*, where each node represents a trade-off, and the expansion of the tree represents all the possible combinations. Each time a node is appended, consistency of the tree must be checked [12]. Thus, the tree might grow large. This structure alone does not bring any benefits with respect to the basic algorithm, but it enables some optimization. First, the tree includes much redundant information which is possible to remove: it is possible to remove a trade-off chain which are *subsumed* by another trade-off chain. The tree is used after the regular skyline is computed, by implementing it in a traditional algorithm (e.g., BNL): for each two objects is checked whether there exists a trade-off dominated by the first and dominating the second; in this case, the latter object is removed.

In conclusion, trade-off skylines allow to customize the output according to user's preferences but, although they are a subset of traditional skylines, they do not guarantee control over size of the output.

# 7 ORD and ORU Queries

ORD and ORU [10] are the last operators we analyze in this survey: considering the previous ones, which typically lose some benefits of top-k or skyline queries, these techniques maintain both user personalization over attributes and strict control of output size, making them, from this point of view, the best overall package among all the others. They guarantee the output-size specification (OSS), personalization and relaxed preference input, i.e., user preference is not considered a strict bound but rather an estimation: indeed, these approaches expand incrementally the weights of preference in all directions, meaning that they deal with a hypersphere and not a polytope as in the other cases. In the original weights $w$, we are dealing with top-$k$ queries, while increasing the radius means shifting towards *domination* (i.e., skylines). The stopping radius is determined by the specified size of the output. ORD operator stresses OSS property, relaxed-input, and dominance, while ORU is more related to the concept of ranking by utility.

An important concept related to the expansion radius $\rho$ is *$\rho$-dominance*. Given the preference vector $v$ and $w$ the best-effort estimate of the user's preference vector, called seed, we consider the vector within distance $\rho$, i.e., $|v - w| \leq \rho$: record $a$ *$\rho$-dominates* record $b$ if $a$ scores as list as high as for every vector $v$ strictly higher for at least one. Records dominated by fewer than k other constitute the *$\rho$-skyband*.

These two techniques provide a different approach with respect to the ones previously described in this survey, and their implementation deliver a practical and scalable performance.

## 7.1 ORD

The ORD operator outputs, given a seed vector $w$ and output size $m$, the ρ-skyband for the smallest $\rho$ that produces $m$ records. The computation of the ρ-skyband set is based on the concept of *inflection radius*: given a record $r$, its inflection radius is defined as the smallest $\rho$ where r is dominated by less than k records. The basic approach consists of computing first the k-skyband set, and for each record, compute the inflection radius: the output represents the m records with the smallest inflection radii. However, computing the entire k-skyband at the beginning is expensive, and there is a more efficient way to solve the problem.

|   | Price | Mileage | Seed score |
|---|---|---|---|
| b | 2 | 10 | 16 |
| c | 3 | 11 | 20 |
| d | 4 | 14 | 26 |
| e | 5 | 13 | 28 |
| g | 6 | 9 | 27 |
| h | 5,7 | 12 | 29,1 |
| f | 5 | 7 | 22 |
| a | 1 | 12 | 15 |

Table 3: Motorcycle dataset example (units in thousands).

Consider our running example and a seed vector $w$ = [3, 1]. Table 3 shows each point in the dataset and its associated score evaluated over the seed vector. The non-dominated tuples which fall in the third category are *a*, *b*, and *f*: *b* and *f* score higher than *a*, but, nonetheless, none of them dominates *a*. Regarding tuple *a*, we can calculate its inflection radius, given k = 2. For ρ=1, by adding or subtracting to each dimension the ρ value we can compute the *v* vectors (e.g., [3,1]+[-1,0]=[2,1]), which are [2,1] [3,0] [3,2] [4,1]: in this case, after the computation of the new scores for each vector, we see that no tuple within *b* and *f* scores higher than *a*; so, we proceed by expanding the radius. For ρ=2, one of the *v* vectors is [1,1]: in this case, we can say that *a* is ρ-dominated by less than two tuples, as the new values of *a*, *b*, *f* are, respectively, 13, 12, 13.

A better ORD algorithm invokes a progressive k-skyband retrieval that places records one by one in a candidate set, by fetching them in a decreasing order for *w*. This is important because it is possible to compute the inflection radius of each candidate on the fly, given that only k-skyband records with higher score (which are already fetched) may ρ-dominate the considered record. The algorithm continues fetching the candidates until the set reaches size m+1. At this point, the candidate with largest inflection radius is removed. So, the algorithm keeps track of the largest inflection radius ρ̲ of the record previously removed, and starts fetching ρ̲-skyband records, in decreasing order. At this point new, as new candidates are fetched, new evictions are made and ρ̲ keeps shrinking to have m candidates in the set. The algorithm terminates when there are no more records to be fetched, and it becomes more selective at each step.

## 7.2 ORU

The second operator, ORU, is strictly related to ranked by utility: given a seed vector w and output size m, it returns the records that belong to top-k for at least one preference vector within radius ρ. The computation of this operator involves the computation of the *convex hull*, which is the smallest convex polytope that encloses all the records of a dataset. The algorithm used to compute ORU is precomputation-free. The first step requires to produce an overestimate of ρ which guarantees an output size of at least *m* (this can be done by computing the upper hull, i.e., the set of facets that have normal directed to the positive section of the plane and find the m top regions with smallest mindist to w and use the largest mindist among them as ρ̲). With an incremental ρ-skyline algorithm, the set is extended for the immediately larger radius around w that admits exactly one new record. This algorithm goes on until the resulting set comprises *m* records. Afterwards, the algorithm computes the upper hull of each record, until it gets a layer with *m* records. The final radius reported by the ρ-skyline algorithm is named ρ̲. With ρ̲ the algorithm computer the ρ-skyband for the actual k specified size, with a standard k-skyband algorithm. The obtained ρ̲-skyband is a superset of the ORU, and so the records are placed in a candidate set: gradually expanding ρ, starting from zero, towards ρ̲ to progressively output confirmed records.

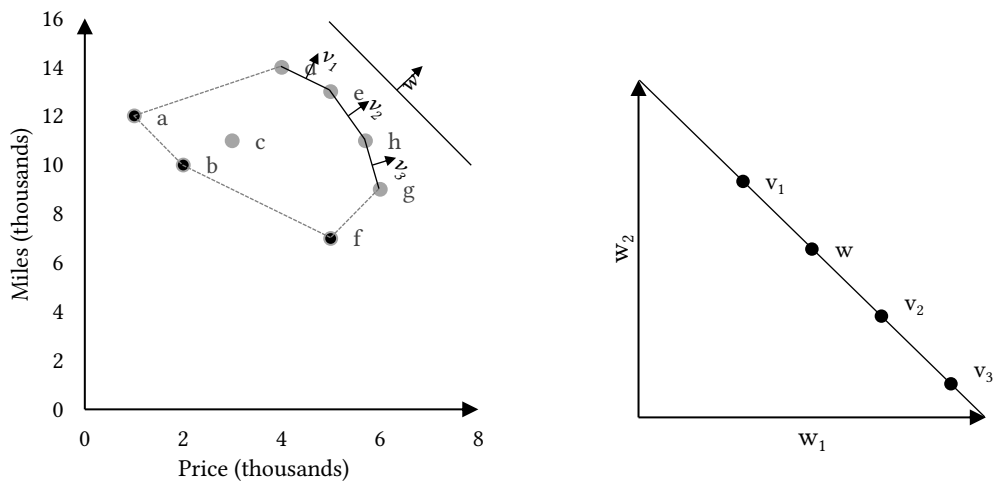

Figure 7: running example convex hull and top record

Example in Figure 7 shows the convex hull polytope (dashed lines), which comprises all data. The upper hull is highlighted in black continuous line. The upper hull is composed by the following facets, i.e., *d-e*, *e-h* and *h-g* (their norms are, respectively, $v_1$, $v_2$, $v_3$). For the *w* vector shown in figure, the top record is *e*, i.e., the first record met by the hyper-plane, normal to *w*, that sweeps from the top corner to the origin. Figure 7 (right) shows the preference region: each segment represents the top-region of the upper hull points; for instance, the top-region of point *d* is the segment *1-$v_1$*, the one of point *e* is the segment $v_1$- $v_2$ and so on. Given a specific $\rho$ (the distance starting from *w*), the best records (the highest scoring ones) will belong to the adjacent set, i.e., the set {*d, h*}.

# 8 Comparison

In this survey we discussed a general overview of five interesting techniques designed to surmount the shortcomings of top-k and skyline queries. Each of these algorithms face the problem by applying different mathematical foundations and theoretical backgrounds. This diversity among the approaches leads to very different results, in most cases not completely satisfactory but still providing great benefits over traditional techniques in terms of customization and flexibility.

|  | Control of cardinality | Overall view of interesting results | Input flexibility |
|---|---|---|---|
| Ranking Queries | Yes | No | No |
| Skyline Queries | No | Yes | Yes |
| ε-Skyline | Partial | Yes | Yes |
| UTK | Partial | Yes | Yes |
| R-Skyline | Partial | Yes | Yes |
| k-Regret | Yes | Yes | No |
| Trade-off | Partial | Yes | No |
| ORD & ORU | Yes | Yes | Yes |

Table 4: comparison summary.

The most refined and capable operators are ORD and ORU, which grant strict control over cardinality of results (as for top-k), input customization (as they make use of the utility function) but still leading to an overall view of the results. Also, k-regret queries are a good option if control over cardinality is strictly required in a particular setting. Although these operators might seem the best overall package, in some applications it might be useful to retrieve a set which gives a wider view of the dataset to the user. In this scenario, ε-skyline, F-skyline, UTK and trade-off queries should be considered. They provide as result a smaller (for ε-skyline also larger) set with respect to traditional skylines. However, result size might vary, as they indirectly control result's cardinality with some parameters: this might lead to difficult implementation as fine-tuning them is challenging. None of these operators provide a completely user agnostic analysis of the dataset, like normal skylines do.

Five of the techniques we have seen, i.e., F-skyline, ε-skyline, trade-off skyline, UTK and ORD/ORU are based on the notion of dominance: each of them adds restrictions to achieve customization and cardinality control. $\mathcal{F}$-dominance, ε-dominance, trade-off dominance, r-dominance and ρ-dominance are the definitions on which the resulting sets are evaluated. As we have seen, the most intuitive one is ε-dominance, which adjusts the dominance region by assigning a value to the ε parameter: when ε = 0, the result is the same of the traditional skyline, but assigning other values generates larger or smaller sets;

this means that ε-skyline is the only technique able to retrieve a resulting set greater than normal skylines. Similarly, $\mathcal{F}$-dominance modifies the dominance region, with the support of families of functions, making it smaller than the one obtained with traditional skylines. In both cases, the size of the result is not predictable. The particularity of R-dominance is related to the fact that it is derived by relaxing the weights of the preference vector. $\mathcal{F}$-dominance and R-dominance are strictly related, as the techniques based on them are one (F-skyline) a particular case of the other (UTK): basically, UTK are capable of evaluating tuples $\mathcal{F}$-dominated by less than $k$ others and thus, extending F-skyline to the case of k>1. Differently, trade-off dominance is related to the use of a scoring function (this can be viewed as special case of F-skyline, where the family of functions is the sum of weights) and the peculiarity of this approach is the focus on user preferences. The last definition, ρ-dominance, is close to $\mathcal{F}$-dominance as both techniques consider the score of tuples given a family of preference input; nonetheless, ρ-dominance adds a strict control over the cardinality of the resulting set, and this feature makes it the only technique capable of full control over output size. The only technique not directly based on dominance is k-regret query, which relies on the concept of happiness (regret) of the user: minimizing the regret ratio is the key criterion of this paradigm. Furthermore, k-regret queries do not require any utility functions produced by users, but still they maintain a controllable output size (exactly k tuples are returned). For these reasons, k-regret queries might seem the superior technique among the others.

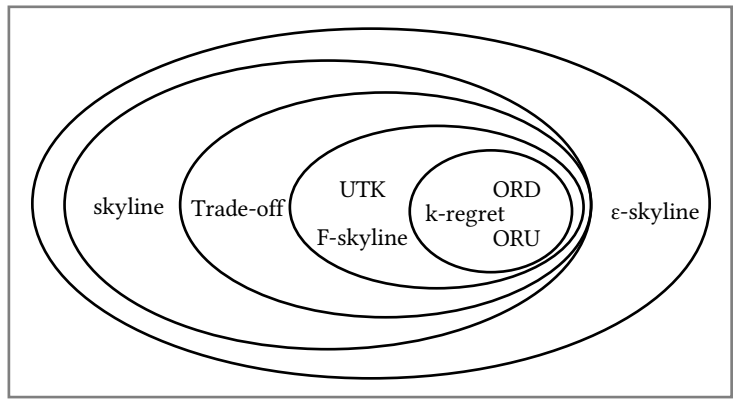

Figure 7: relation between the different techniques regarding their output size.

Another aspect we want to underline is the relationship between F-skylines and ORD/ORU operators. F-skylines return all the tuples that rank as top-1 for any weight combination, given a set of linear constraints on the weights. ORD and ORU represent the most refined extension of this approach, as they are capable of computing top-k (with k>1) results, by an incremental exploration of the preference region.

The complexity of implementation of the analyzed operators is strictly dependent on the parameter they use to control the output size: in fact, the operators that use a scoring function have the possibility to mine the weight instead of having them provided by the users. Trade-off queries still require the user to declare their preference but, as opposed to top-k queries, they do not require specific weights on the attributes: the user should only declare some preference values over a subset of the attributes.

## 9 Conclusion

Bridging the gap between top-k and skyline queries is still an open debate among databases scholars. In this survey we proposed and analyzed the most relevant techniques developed in the past recent years and we clarified the state of the art in this science, by describing the fundamentals behind these algorithms and their working mechanisms. For each of these techniques we analyzed their main pros and cons, also by comparing them with the traditional top-k and skyline approaches. The reader should now have a clear overview regarding these topics and the means to deepen their studies.